\documentclass[a4paper,10pt]{article}
\usepackage{graphics,epsfig}
\usepackage{amssymb}

\begin{document}

\title{Axionic Shortcuts for High Energy Photons
}

\author{\Large{A. Nicolaidis} \medskip \\
Theoretical Physics Department \\
University of Thessaloniki \\
54124 Thessaloniki, Greece \\
nicolaid@auth.gr}
\date{\today}
\maketitle

\begin{abstract}
We study the photon axion mixing in the presence of large extra
dimensions. The eigenvalues and eigenstates of the mixing matrix
are analyzed and we establish the resonance condition for the
total conversion of a high energy photon into a Kaluza-Klein (KK)
axion state. This resonant transition, a photon transformed into a
KK axion travelling freely through the bulk and converting back
into a photon, may provide a plausible explanation for the
transparency of the universe to energetic photons. If the brane we
live in is curved, then there are shortcuts through the bulk,
which the axion can take. Within our model, the photons having the
appropriate resonance energy are using the axionic shortcut and
arrive earlier compared to the photons which follow the geodesic
on the brane. We suggest that such axionic shortcuts are at the
root of the dispersion of time arrival of photons observed by the
MAGIC telescope. We indicate also the cosmological significance of
the existence of axionic shortcuts for the photon.
\end{abstract}

\subsubsection*{Introduction}
\label{sec:intro}
The main motivation for introducing new physics comes from the need to provide a unified theory in which two disparate scales,
the electroweak scale ($M_W \sim 200 GeV$) and the Planck scale ($M_{Pl} \sim 10^{19} GeV$), can coexist. A novel approach has been suggested
to alleviate the hierarchy problem \cite{1}. Our four-dimensional world is embedded in a higher dimensional space with
D dimensions ($D = 4+\delta$). While the standard model fields are constrained to live on the 4-dimensional brane, gravity can freely
propagate in the higher-dimensional space (bulk). The fundamental scale $M_f$ of gravity in D dimensions is related to the
observed 4-dimensional Planck scale $M_{Pl}$ by
\begin{equation}
 M_{Pl}^2 = M_f^{2+\delta} V_{\delta}
\end{equation}
where $V_{\delta}$ is the volume of the extra space. For a torus configuration
\begin{equation}
 V_{\delta} = (2 \pi)^{\delta} R_1 R_2 \cdots R_{\delta}
\end{equation}
with $R_i$ ($i=1, 2, \cdots ,\delta$) the radii of extra dimensions. Then for a sufficiently large volume $V_{\delta}$ the fundamental
scale of gravity $M_f$ can become as low as $M_W$. In this radical way the hierarchy problem ceases to exist as such.

The prospect of gravity becoming strong at TeV energies, opens the possibility of studying gravity in particle collisions at accessible
energies (at present or in the near future). To that respect, salient features of the cosmic ray spectrum (the "knee") have
been attributed to gravitational bremsstrablung \cite{2}. By reproducing the cosmic ray sprectum, the parameters of the low scale
gravity can be inferred ($\delta \sim 4$ and $M_{f} \sim 8 TeV$). Within the same scenario, we may anticipate the formation of black holes
in high energy collisions, signaled by an enhancement of backward scattering, associated with rainbow-like diffraction patterns \cite{3}.
Besides the graviton, fields which are standard-model singlets, like a sterile neutrino \cite{4,5,6} or an axion \cite{7,8} can freely
propagate in the bulk. These particles accrue then an infinite tower of Kaluza-Klein (KK) excitations and the issue of mass eigenstates
 and mixing has to be revisited.

In the next part we consider the photon-axion mixing, induced by a magnetic field, in the context of large extra dimensions.
Compared to the usual oscillations, novel features appear linked to the presence of the new scale, the radius of the extra dimension
 (or its inverse, the mass of the Kaluza-Klein excitation). We study in detail the eigenvalues and eigenstates of the mixing matrix
and establish the resonance condition for the total conversion of a high energy photon into a KK axion state. In the third part we examine
the astrophysical implications of our formalism, notably the production and propagation of photons in sites of strong magnetic
 fields (neutron stars, GRBs). A photon-generated KK axion can take a "shortcut" through the bulk and appear earlier,
compared to a photon traveling along a geodesic on the brane. We quantify this effect and analyze its relevance for the timing
of photons observed by the MAGIC telescope during an activity of Markarian 501. The axionic shortcuts broaden also the photon
light cone, providing correlations on superhorizon scales and enhancing the uniform appearance of the universe. At the end we present
 our conclusions.

\subsubsection*{Axion in large extra dimensions}
The Peccei-Quinn (PQ) solution to the strong CP problem in QCD
\cite{9}, predicts the existence of a neutral, spin-zero
pseudoscalar
 particle, the axion. Axions and photons oscillate into each other in an external magnetic field \cite{10,11} due to the interaction term
\begin{equation}
 \mathcal{L}_{int} = \frac{1}{f_{PQ}} a F_{\mu\nu} \tilde{F}_{\mu\nu} = \frac{4}{f_{PQ}} a \vec{E} \cdot \vec{B}
\end{equation}
where $F_{\mu\nu}$ is the electromagnetic field tensor, $\tilde{F}_{\mu\nu}$ is its dual, $a$ is the axion field. The mass $m_{PQ}$
 of the standard axion, as well as the axion-photon coupling are inversely proportional to the scale $f_{PQ}$.
For a recent account of the axion searches see ref 12.

To simplify the calculation for the higher-dimensional case, we consider one extra compact dimension y and a singlet axion field
$a(x^{\mu},y)$. Projected into the brane the axion field will appear as a collection of KK modes $a_{n}(x^{\mu})$, each having
a mass $m_n = \frac{n}{R}$, where $R$ is the compactification radius. The coupling of the KK axions to the photon is universal
 \cite{7,8,13,14}
\begin{equation}
 \mathcal{L}_{int} = \frac{1}{f_{PQ}} \sum_n a_n F_{\mu\nu} \tilde{F}_{\mu\nu}
\end{equation}
 The mixing matrix M between the photon state $A_{||}$ parallel to the magnetic field B, the standard PQ axion $a_0$ and the
 KK axions $a_n$ is \cite{8}
\begin{equation}
 \mathbf{M} = \left(
  \begin{array}{ccccc}
   \Delta_{\gamma} & \Delta_{B} & \Delta_{B} & \ldots & \Delta_{B}\\
   \Delta_{B} & \Delta_{0} & 0 & \ldots & 0 \\
   \Delta_{B} & 0 & \Delta_{1} & \ldots & 0 \\
   \vdots & \vdots & \vdots & \ddots & \vdots \\
   \Delta_{B} & 0 & 0 & & \Delta_{N}
  \end{array} \right)
\end{equation}
where
\begin{equation}
 \Delta_{\gamma}=\frac{\omega_{pl}^2}{2E}, \quad \Delta_{0}=\frac{m_{PQ}^2}{2E}, \quad
\Delta_{n}=\frac{n^2}{2ER^2}, \quad \Delta_{B}=\frac{4B}{f_{PQ}}
\end{equation}
 The plasma frequency is $\omega_{pl}^2 = (4 \pi a n_e)/m_e$ for an electron density $n_e$.

The eigenvalues $\lambda$ of the mixing matrix M obey the equation
\begin{equation}
 \Delta_{B}^2 \left[ \frac{1}{\lambda -\Delta_0} + \frac{1}{2 \lambda} \left( xcot x -1\right)\right]= \lambda - \Delta_{\gamma} \label{7}
\end{equation}
 with
\begin{equation}
 x^2 = 2 \pi^2 E R^2 \lambda \label{8}
\end{equation}
In the limit $R\rightarrow 0$, the KK modes become unaccessible
and eqn.(\ref{7}) is reduced to
\begin{equation}
 \Delta_{B}^2 = (\lambda - \Delta_{0}) (\lambda - \Delta_{\gamma})
\label{9}
\end{equation}
describing the ordinary two-level mixing. For a relatively small radius, we may use the Taylor expansion $xcot x -1 \simeq - \frac{x^2}{3}$
 and we obtain an equation similar to eqn.(\ref{9}), with $\Delta_{\gamma}$ replaced by an effective $\Delta_{eff}$, given by
\begin{equation}
 \Delta_{eff} = \Delta_{\gamma} - \frac{\pi^2}{3} E R^2 \Delta_{B}^2
\label{10}
\end{equation}
In the absence of a magnetic field the unperturbed eigenvalues are
the diagonal elements  $\Delta_{\gamma}$, $\Delta_{0}$, $\Delta_{1}$,
$\cdots$ ,$\Delta_{N}$. In the presence of a strong magnetic field,
the eigenvalues $\lambda_{n}$ are approximately given by
\begin{equation}
 \lambda_{n} \simeq \frac{(n + \frac{1}{2})^2}{2 E R^2}
\label{11}
\end{equation}
The eigenstate corresponding to the eigenvalue $\lambda_{n}$ is defined by the row vector $E_n = (E_{n \gamma}, E_{n0}, E_{n1},\cdots ,E_{nk},\cdots ,E_{nN})$.
The elements of the eigenvector satisfy the equation
\begin{equation}
 E_{nk} = \frac{\Delta_{B}}{(\lambda_n - \Delta_k)} E_{n \gamma}
\label{12}
\end{equation}
The amplitude for a photon converted into the nth KK axion and then back to photon is given by $|E_{n \gamma}|^2$. This amplitude
is maximized whenever the resonance condition
\begin{equation}
 \lambda_n = \frac{1}{2} \{(\Delta_{0} + \Delta_{\gamma}) \pm \left[ (\Delta_{0} - \Delta_{\gamma})^2 + 4\Delta_B^2\right]^{\frac{1}{2}}\}
\label{13}
\end{equation}
is satisfied. For $n = 1$ the resonance is a narrow one and the amplitude $A ( \gamma \rightarrow \gamma)$ is enhanced. For
 large n, the resonance is a broad one, many KK states interfere destructively and the amplitude $A ( \gamma \rightarrow \gamma)$
is reduced.

\subsubsection*{Astrophysical implications}
There are important constraints on the axion system and the size of the extra dimension. The $f_{PQ}$ scale has to be larger than
$10^{11} GeV$, with a corresponding mass for the axion $m_{PQ} \leqslant 10^{-2} eV$ \cite{14}. For the compactification radius R
 we consider the range $10^{-1} cm \leq R $, respecting the available experimental data. We are interested in high
 energy photons, produced in sites where strong magnetic fields reign. Under these terms, the resonance condition, eqn. (\ref{13}),
 is simplified for $n=1$ to
\begin{equation}
 \frac{1}{2 E R^2} = \Delta_{B}
\label{14}
\end{equation}
 or in actual units
\begin{equation}
 \left( \frac{E}{500 GeV}\right) \left( \frac{R}{10^{-3}cm}\right)^2 \left( \frac{B}{10^{7} G}\right) \left(\frac{10^{12} GeV}{f_{PQ}}\right)=1.0
\label{15}
\end{equation}
Within this scheme, photons of high energy produced in an active nucleus are transformed through an MSW type resonance into KK axions,
 which travel unimpeded before being reconverted back into photons. For photons not satisfying the resonance condition,
 the probability to transit as KK axions is reduced and therefore these photons are limited by the opacity of two photon
annihilation into an electro-positron pair. Our approach offers a
plausible explanation for the transparency of the universe to
energetic photons \cite{15}. Photons of the appropriate energy,
are transformed into KK axions, travel freely in the bulk space,
before returning back into the brane and observed again as
photons. The same mechanism may provide also high energy photons
 escaping the GZK cutoff \cite{16}.

The advent of imaging atmospheric Cerenkov telescope like HESS,
MAGIC, VERITAS, FERMI, CTA, allows the detection of photons from
astrophysical sources (neutron stars, GRB, AGN) in the high energy
window from $100 GeV$ to few $TeV$. The MAGIC telescope
 analyzed the timing of photons originating from the Mkn 501 source \cite{17}. It was found that the photons in the $0.25 - 0.6 TeV$
 energy range precede by 4 minutes the photons in the $1.2 - 10 TeV$ energy band. The observed features might be possible to be
explained within an astrophysical context \cite{18}. A particle physics solution, within the framework of quantum gravity
\cite{19,20} has been suggested also. Quantum fluctuations of space-time lead to a speed of light dependent upon energy, thus creating
 a dispersion in the time arrival of photons \cite{21}.

The transition of a photon to a KK axion, offers another alternative to analyze the observed dispersion in time arrival of high
 energy photons. A particle, travelling from a point on the brane to another point on the brane, may take a "shortcut" by
 following a geodesic in the bulk and arriving earlier compared to a particle which follows a geodesic on the brane \cite{22}. In our
 case a photon, transformed into a KK axion traveling through the bulk, may reappear earlier on the brane, compared to a photon stuck in the brane.
 There are a number of ways shortcuts emerge in theories with extra dimensions. In a brane containing matter and energy, self-gravity
 will induce a curvature to the brane, so that the brane becomes concave towards the bulk in the null direction. Then we can find
geodesics in the bulk propagating signals faster compared to the
geodesics in the brane \cite{22}. A phenomenology, with a sterile
neutrino shortcuting through the bulk has been developed \cite{23,
24}, accounting for all neutrino oscillation data.
 A $1+1$
dimensional toy
 model may exhibit the expected behavior \cite{23}. In a Minkowski metric
\begin{equation}
 ds^2 = dt^2 -dx_1^2 -dx_2^2
\end{equation}
the curved brane is represented by
\begin{equation}
 x_2 = A sinkx_1
\end{equation}
while the bulk geodesic is given by $x_2 =0$. A signal transmitted
through the bulk will appear as having a superluminal speed.
Equivalently a difference in time arrival will be observed given
by
\begin{equation}
 \frac{t_{\gamma}-t_{\alpha}}{t_{\gamma}} \simeq \left( \frac{Ak}{2}\right)^2
\end{equation}
where $t_{\gamma}$ is the time it takes for the photon in the brane and $t_{\alpha}$ is the corresponding time for the photon
 which uses the axionic shortcut. Within our model, the photons having the resonance energy,
 eqn.(\ref{15}), use
the axionic shortcut and arrive earlier compared to the other photons of different energies which propagate along the brane.
>From the MAGIC experimental data we infer that the $0.25-0.6 TeV$ photon energy brackets the aforementioned resonance energy.
 The amount of the time difference is determined by the brane shape-parameter $Ak$ and the rauge of the magnetic field
 B. The magnetic field near the core of a GRB or a blazar reaches
 high values. We adopt the average value of $B=10^{7}$ G. By
 appealing to the Hillas criterion \cite{25} we obtain a range
 for the magnetic field of the order of $10^9$ km. Then the set of
 values $R=10^{-3}$ cm, $f_{PQ}=10^{12}$ GeV, $B=10^{7}$ G, $E=500$ GeV, $\Delta t=4$ min, implies
 for the shape parameter the value $Ak \simeq 1$. Monitoring the astrophysical sites, where high energy photons are
produced in the presence of strong magnetic fields, might be
 revealing. We might observe the disappearance of high energy photons, as a result of a photon - KK axion
transition. Or, we might notice the sudden appearance of high
energy photons, connected to the intrusion of KK axions from the
bulk into the brane.

The presence of a KK axion may affect aspects of current cosmology
\cite{26,27}. It is beyond the scope of the present work a
detailed account of these aspects. It suffices to remark the
following. A KK axion living in the bulk and with a mass
determined as $m \sim \frac{1}{R}$ may assume the role of the dark
matter particle.  Within the evolution history of the universe,
the brane shape-parameter Ak evolves with time. It is anticipated
that in the early Universe the parameter Ak would have a
significantly higher value. If $r_{\gamma}$ is the horizon
 radius for the causal propagation of photon signals on the brane and $r_a$ is the horizon radius for the causal propagation of photon signals mediated by a KK axion through the bulk, then in the early moments of the
Universe the ratio $r_{\alpha}/ r_{\gamma}$ may attain high values
of the order $10^3$ \cite{28}. The communication of information
 over length scales $r_{\alpha} >> r_{\gamma}$ might lessen the severity of the horizon problem and change the initial
conditions for inflation. Photon correlations on superhorizon scales lead also to a redefinition of the last scattering
surface to earlier times.

\subsubsection*{Conclusions}
The idea that our four-dimensional world is embedded in a higher dimensional space has triggered an outburst of creative and profound
 research in particle physics and cosmology. A variety of models have been developed where gravity and gravitational phenomena are
observable at accessible energies. Besides the graviton, living in
the bulk, particles which are singlets under the standard model
may live in extra dimensions.  In the present work we studied the
photon-axion oscillation in the presence of extra dimensions.
 Next to the Peccei-Quinn scale $f_{PQ}$ and the magnitude of the magnetic field B, a new scale is introduced, the size R of the extra
 dimension, or its inverse the KK mass excitation. An MSW-type resonance occurs at high energies between the photon and the KK axion.
 This resonance transition A ($\gamma \rightarrow \alpha \rightarrow \gamma$)
allows the photon to travel unimpeded from a dense and opaque
medium, offering the ground for a rich phenomenology. It is
 highly plausible that our brane is curved creating the possibility of shortcut through the bulk. We suggested that an axionic
shortcut may be at the origin of the dispersion in the time
arrival of the photons observed by the MAGIC telescope. Our model
provides distinct signatures, relatively easy to verify or
disprove. The photons arriving ahead of time, or the photons
appearing or disappearing suddenly,  satisfy the
 resonance condition eqn.(\ref{15}), and therefore the energy E of these photons times the magnetic field B of the astrophysical
 site where they are produced is a constant.  This behavior should be contrasted to the expectations derived
within the spacetime foam mode \cite{21}, where the time delay of
photons steadily increases with energy.The photon telescopes,
under operation or construction, through a fine binning of the
photon energy may unravel the underlying mechanism.  We indicated
also the cosmological significance of an axionic shortcut for the
photon.
 It appears that the topic of axionic shortcuts for high energy photons is a highly interesting one, inviting theoretical insights and
 experimental investigation.

~

\textbf{Acknowledgements} The present work was presented during a
seminar at the Complutense University, Madrid, within the
scientific exchange program between Greece and Spain. I would like
to thank Prof. Marina Ramon Medrano for the warm hospitality and
Prof. Eckart Lorenz, Prof. Jose Luis Contreras for stimulating
discussions.

\end{document}